\documentstyle[12pt]{article}
\newcommand{\cl}{\centerline}
\def\beq{\begin{equation}}
\def\eeq{\end{equation}}
\def\bea{\begin{eqnarray}}
\def\eea{\end{eqnarray}}
\def\bq{\begin{quote}}
\def\eq{\end{quote}}

\def\PLB{{\it Phys. Lett.} }
\def\PRL{{\it Phys. Rev. Lett.} }
\def\NP{{\it Nucl. Phys.} }
\def\PR{{\it Phys. Rev.} }

\def\IJMP{{\it Int. J. Mod. Phys.} }

\def\SNP{{\it Sov. J. Nucl. Phys.} }

\def\gappeq{\mathrel{\rlap {\raise.5ex\hbox{$>$}}
{\lower.5ex\hbox{$\sim$}}}}

\def\lappeq{\mathrel{\rlap{\raise.5ex\hbox{$<$}}
{\lower.5ex\hbox{$\sim$}}}}

\begin{document}
\topmargin -2.0cm
\oddsidemargin -0.3cm
\evensidemargin -0.8cm
\pagestyle{empty}
\begin{flushright}
{CCUTH-95-04\\ IP-ASTP-14-95}
\end{flushright}
\vspace*{5mm}
\begin{center}
{\bf PQCD ANALYSIS OF INCLUSIVE SEMILEPTONIC DECAYS OF $B$ MESONS}\\
\vskip 1.0cm
\cl{Hsiang-nan Li$^1$\ and Hoi-Lai Yu$^2$, }
\vskip 0.3cm
\cl{$^1$Department of Physics, National Chung-Cheng University,}
\cl{Chia-Yi, Taiwan, R.O.C.}
\vskip 0.3cm
\cl{$^2$Institute of Physics, Academia Sinica, Taipei, Taiwan, R.O.C.}\par

\vspace*{0.75cm}
{\it \today}\\
\vspace*{0.5cm}

\vspace*{1.0cm}
{\bf Abstract}
\vskip 0.3cm
\end{center}

We develop the perturbative QCD formalism for inclusive semileptonic
$B$ meson decays, which includes Sudakov suppression from the resummation
of large  radiative  corrections  near  the high end of charged
lepton energy.  Transverse  degrees of freedom of partons are introduced to
facilitate the factorization  of $B$ meson decays.  Ambiguities appearing
in the quark-level analysis  are then avoided.  A universal distribution
function, arising from the nonperturbative Fermi motion of the $b$ quark,
is constructed according to the heavy quark effective field theory
based operator product expansion, through which the mean and the width of
the distribution function are related to hadronic matrix elements of local
operators. Charged lepton spectra of the $B\to X_u\ell\nu$ decay are
presented. We find 50\% suppression near the end point of the spectrum.
The overall suppression on the total decay rate is 8\% for the free quark
model, and is less than 7\% for the use of smooth distribution functions.
With our predictions, it is then possible to extract the
Cabibbo-Kobayashi-Maskawa matrix element $|V_{ub}|$ from experimental data.
We also discuss possible implications of our analysis when confronted with
the rather small observed semileptonic branching ratio in $B$ meson decays.

\newpage
\setcounter{page}{1}
\pagestyle{plain}

\section{Introduction}
\hskip 0.6cm
The studies of semileptonic decays in heavy mesons within the framework
of perturbative quantum chromodynamics (PQCD) dated back to the seventies
\cite{aa}. The mass $M_Q$ of a heavy quark $Q$ provides a rationale of this
approach. The advantage of the PQCD formalism is that it provides a natural
normalization of decay amplitudes. This point is of great importance to
obtain a model independent extraction of the Cabibbo-Kobayashi-Maskawa
(CKM) matrix elements, which are the key phenomenological parameters in
understanding the symmetry breaking physics of the standard model. We have
shown that PQCD is applicable to the exclusive  $B\to \pi \ell\nu $ and
$B\to \rho \ell \nu $ decays \cite{bb}, and an upper limit of $|V_{ub}|$
around 2.7 $\times$ 10$^{-3}$ was extracted from experimental data \cite{cc}
directly. Recently, there have been controversies concerning the inclusive
semileptonic branching ratio of $B$ meson decays \cite{af,bf}. Even new
physics was proposed \cite{af} to resolve the discrepancy between theoretical
predictions and experimental outcomes. It has been argued that PQCD radiative
corrections may play an important role in the semileptonic decays $B\to
X\ell\nu$ \cite{bf}. In view of this, it is essential to first sort out the
correct PQCD contributions to these decays.

In the pioneering works of Chay {\it et al.} \cite{ee} and Shifman
{\it et al.} \cite{dd} a systematic expansion of relevant hadronic matrix
elements in semileptonic decays in the inverse power of $M_Q$ was obtained
by combining heavy quark effective field theory (HQEFT) and the method of
operator product expansion (OPE). It was shown that the differential decay
rate $d\Gamma/dq^2dE_\ell$ can be consistently calculated, when it is
suitably averaged over the charged lepton energy $E_\ell$. Here
$q^2=(p_\ell+p_\nu)^2$ is the lepton pair invariant mass with $p_\ell$ and
$p_\nu$ the charged lepton and neutrino momenta, respectively. At leading
order, the expansion reproduces the naive parton model predictions \cite{ff}
of free heavy quark decays. Next-to-leading order corrections, starting
at ${\cal O}(\Lambda_{\rm QCD}^2/M_Q^2)$, are expected to be small in
heavy meson decays. This approach has also been extended to the case of
nonleptonic decays \cite{gg}.

To extract $V_{ub}$ from the $B\to X_u\ell\nu$ decay, one needs to
measure the charged lepton spectrum near the high end of $E_\ell$,
such that the huge $B\to X_c\ell\nu$ background stops contributing. The
measurement must be performed within an accuracy of several hundred MeV,
because the energy difference between the end points of the $b \to c$ and
$b\to u$ transitions is only about 330 MeV. Unfortunately,
OPE, the theoretical tool, breaks down in this region. Since the expansion
parameter $1/M_Q$ should be replaced by $1/(M_Q - q\cdot v)$ when $q$ is
not small at the end point, OPE is not reliable.
The heavy quark velocity $v$ is defined by
$P_Q=M_Q v$, $P_Q$ being the heavy quark momentum. To circumvent the
difficulty, Neubert \cite{N} and Bigi {\it et al.} \cite{bsuv} have
performed a resummation of OPE, that results in a model independent
``shape function" in the description of the charged lepton spectrum. This
universal shape function can be determined in principle by an infinite
tower of nonperturbative matrix elements expanded in the increasing power
of $1/M_Q$, and has been employed in the study of the inclusive rare
decay $B\to X_s\gamma$ \cite{dsu,neubert}. One can therefore measure this shape
function, say, in the $B \to X_s\gamma$ decay, and then apply it to the
$B\to X_u\ell\nu$ decay to obtain a model independent prediction of
$V_{ub}$. However, under the present experimental situation, one has to
make an ansatz for this shape function according to some QCD constraints.
Data of the decay $B\to X_s\gamma$ will definitely
remove this ambiguity in the choice of shape functions in the future.

On the parton model side, Bareiss and Paschos \cite{kk} argued that the
phase space for the decays is dominated by distances near the light cone,
and one can use the distribution function of the $Q$ quark in the heavy
meson $H_Q$ in the infinite momentum frame to fill up the gap between the
simple heavy quark kinematics and the heavy meson kinematics. This
distribution function can be obtained by measuring the $Q$-quark fragmentation
function in the heavy meson production from $e^+ e^-$ annihilation.
To remove the singularities near the end point of the charged lepton
spectrum, which are due to soft gluon bremsstrahlung, these authors simply
resummed soft gluon corrections in the naive leading (double) logarithmic
approximation and obtained a Sudakov suppression factor. To fill up
the kinematic gap, Altarelli and Petrarca \cite{jj} regarded the light quark
inside $H_Q$ as a quasi-free particle but with a Gaussian spectrum
of Fermi-momentum $p$,
\begin{equation}
\ f(p)=\frac{4p^2}{\sqrt\pi p_f^3} \exp(-\frac{p^2}{p_f^2})\; ,
\label{1}
\end{equation}
where $p_f$ is a free parameter that can be fixed by heavy quark symmetry.
To smooth out the end-point singularities, these authors also resummed
leading soft gluon contributions into a Sudakov form factor.

Korchemsky and Sterman \cite{ks} gave the first PQCD treatment of the
decays $B\to X_s\gamma$ and $B\to X_u\ell\nu$, in which higher-order
corrections were factorized into a soft distribution function, a
jet function and a hard scattering amplitude according to the kinematic
regions of loop momenta. The equivalence between the shape function
\cite{N,bsuv,neubert} and the distribution function was pointed out. Soft gluon
corrections, which correspond to the nonperturbative origin of the
distribution function, were resummed systematically up to next-to-leading
logarithms using the Wilson-loop formalism. However, their analysis is
appropriate only in the end-point region, and the effects of resummation
were not estimated.

In all the above approaches, the factorization was formulated at the
quark-level kinematics, and the missing states inside the kinematic window
between $M_Q$ and $M_H$, $M_H$ being the $H_Q$ meson mass, were populated by
introducing extra heavy quark Fermi motion, which arises from the recoil of
the light partons in $H_Q$. Our approach is formulated at the meson-level
kinematics directly, in which the heavy quark $Q$, carrying a fraction of
the heavy meson momentum, has an invariant mass close to $M_Q$. Our
formalism therefore removes the ambiguity in the definition of the heavy
quark mass $M_Q$, and the kinematic gap is filled up naturally.

In this paper we shall derive the PQCD factorization formula for the
semileptonic decay $B\to X_u\ell\nu$ in a rigorous way, which is suitable
for the entire range of the spectrum. The factorization procedures demand
the inclusion of the transverse degrees of freedom of partons. Hence, we
perform the resummation of large perturbative corrections in the transversal
configuration space using the technique developed in \cite{bb}, which
is also accurate up to next-to-leading logarithms. It can be shown that
our resummation result coincides with that in \cite{ks} at the end point.
The transverse momenta carried by the $b$ quark inside the $B$ meson,
whose distribution is governed by the Sudakov form factor from the
resummation, play an important role here. They diminish the on-shell
probability of the outgoing $u$ quark at the end point, and thus
suppress the singularities.

We define the kinematics of the inclusive semileptonic decays of
heavy mesons in Section 2, and derive the factorization formula for the
charged lepton spectrum, which incorporates the transverse degrees of freedom
of partons. The formula is expressed as the convolution of a hard scattering
amplitude with a jet and a universal soft function. In Section 3, we resum
the large logarithms in these convolution factors by solving a set of
evolution equations. The initial condition of the soft
function is identified as the  distribution function, which is equivalent
to the shape function mentioned above. In Section 4, we construct a
distribution function according to the HQEFT based OPE, and relate the mean
and width of the distribution function to the hadronic matrix elements of
the kinematic operator. Hence, both perturbative higher-order corrections
and nonperturbative corrections from Fermi motion are included in our
formalism. We present numerical results
in Section 5, and show that the Sudakov form factor from the resummation
and the distribution function indeed render the end-point sprectrum
smoother as stressed in \cite{dsu}. Section 6 is the conclusion.

\section{Factorization Theorems}
\hskip 0.6cm
We consider the semileptonic inclusive decays of a $B$ meson,
\begin{equation}
B(P_B) \to \ell (p_\ell) + {\bar\nu}(p_\nu) + {\rm hadrons}\;.
\end{equation}
The three independent kinematic variables are choosen as $E_\ell$, $q^2$
and $q_0$ in our discussion. $E_\ell$ and $q = p_\ell+  p_\nu$ have been
defined in the Introduction, and $q_0$ is the energy of the lepton pair.
With these variables, the triple differential decay rate is written as
\begin{equation}
\ \frac{d^3\Gamma}{dE_\ell dq^2 dq_0}=\frac{1}{256 \pi^4 M_B}|{\cal M}
(E_\ell,q^2, q_0)|^2 \;,
\end{equation}
where $M_B$ is the $B$ meson mass, and the weak matrix element is given by
\begin{equation}
\ {\cal M}=V_{ub} \frac{G_F}{\sqrt 2}\bar\ell \hskip 6pt \Gamma_\mu \nu_\ell
\          \sum_X \langle X|j^\mu|B\rangle \;.
\label{me}
\end{equation}
In eq.~(\ref{me}) $V_{ub}$ is the corresponding CKM matrix element, and
$j_\mu=\bar u \Gamma_\mu b $ is the electroweak current with
$\Gamma_\mu=\gamma_\mu(1-\gamma_5)$.\\

We work in the rest frame of the $B$ meson, and choose the following
light-cone components for relevant momenta,
\begin{eqnarray}
P_B=(P_B^+,P_B^-,{\bf 0_\bot})\;,\;\;\;\;
p_\ell=(p_\ell^+,0,{\bf 0_\bot})\;,\;\;\;\;
p_\nu=(p_\nu^+,p_\nu^-,{\bf p_{\nu\bot}})\;,
\label{niv}
\end{eqnarray}
with $P_B^+=P_B^-=M_B/\sqrt{2}$ and $p_\nu^2=0$.
The independent variables are identified as $p_\ell^+$,
$p_\nu^-$ and $p_\nu^+$, and their relations to $E_\ell$, $q^2$ and $q_0$
are $E_\ell=p_\ell^+/\sqrt{2}$, $q^2=2p_\ell^+ p_\nu^-$ and
$q_0=(p_\ell^++p_\nu^++p_\nu^-)/\sqrt{2}$, respectively. We define
$P_b=P_B-p$ as the $b$ quark momentum, which satisfies $P_b^2
\approx M_b^2$, $M_b$ being the $b$ quark mass. $p$ is the kicks from
the light components inside the $B$ meson, which has a large plus
component $p^+$ and small transverse components ${\bf p_\bot}$. The
purpose of introducing the transverse degrees of freedom will become clear
later. The $b$ quark decays into a $u$ quark with momentum $P_u=P_B-p-q$.
We have distinguished the $B$ meson momentum $P_B$ from the $b$ quark
momentum $P_b$ here.

It is more convenient to employ the scaling variables
\begin{equation}
\ x=\frac{2E_\ell}{M_B}\;,
\ \hskip 0.2cm
\ y=\frac{q^2}{M^2_B} \;,
\ \hskip 0.2cm
\ y_0=\frac{2q_0}{M_B}\;,
\label{a4}
\end{equation}
instead of the dimensionful ones $E_\ell$, $q^2$ and $q_0$.
Note that the scaling variables are defined in terms of the $B$ meson mass
$M_B$, since we formulate the factorization according to the  $B$ meson
kinematics. This differs from the conventional treatment in the literature
\cite{ks}, where the scaling variables were defined in terms of the $b$
quark mass. For massless leptons, it is easy to show, using the momentum
configurations defined in eq.~(\ref{niv}), that the phase space
is given by
\begin{equation}
\ 0\leq x \leq 1,
\hskip 0.2cm
\ 0\leq y \leq x,
\hskip 0.2cm
\ \frac{y}{x}+x\leq y_0\leq y +1\;.
\label{e6}
\end{equation}

In the end point region with $x\to 1$ ($p_\ell^+\to M_B/\sqrt{2}$) and
$y\to 0$ ($p_\nu^-\to 0$), we have $y_0\to 1$ ($p_\nu^+\to 0$) and $p\to 0$.
The $u$ quark then has a large minus component $P^-_u=(1-y/x)
M_B/\sqrt{2}$ but a very small plus component $P^+_u=(1-y_0-y/x)
M_B/\sqrt{2}$, and thus a very small invariant $P^2_u=M^2_B(1-y_0+y)$,
which forms an on-shell jet subprocess. The $u$ quark travels a long
distance of ${\cal O}(1/\Lambda_{\rm QCD})$ before hadronized. Besides,
the $B$ meson is dominated by soft dynamics, which is the origin of the
soft function stated in the Introduction. The remaining
dominant subprocess is the hard one, which contains the weak decay vertices.
Therefore, the important contributions are factorized into the soft $(S)$,
jet $(J)$ and hard $(H)$ subprocesses as shown in fig.~1.

The factorization formula for the inclusive semileptonic decay
$B\to X_u\ell\nu$ is written as
\begin{eqnarray}
\frac{1}{\Gamma^{(0)}_\ell}\frac{d^3\Gamma}{dxdydy_0}
&=&M_B^2\int^{z_{\rm max}}_{z_{\rm min}}{dz} \int d^2{\bf p_\bot}
\nonumber \\
&&\times
S(z,{\bf p_\bot},\mu)J(z,P_u^-,{\bf p_\bot},\mu)
H(z,P_{u}^-,{\bf p_\bot},\mu)\;,
\label{as}
\end{eqnarray}
with the momentum fraction $z$ defined by $z=P^+_b/P^+_B=1-p^+/P^+_B$ and
$\Gamma^{(0)}_\ell=\frac{G_F^2}{16\pi^3}|V_{ub}|^2M^5_B$ \cite{ks}. $\mu$ in
eq.~(\ref{as}) is the renormalization and factorization scale. The triple
differential decay rate is, of course, $\mu$ independent. Note that
in the region $y\to x \sim 1$ the outgoing $u$ quark becomes
soft and eq.~(\ref{as}) fails. We shall show that contributions from
this dangerous region are suppressed by phase space.
The upper limit of $z$ takes the value $z_{\rm max}=1$ in our analysis.
If performing the factorization according to the $b$ quark kinematics,
one must assume $z_{\rm max}=M_B/M_b$, which is greater than 1, in order
to fill up the kinematic window. It has been
explained \cite{ks} that $z_{\rm max}>1$ is not allowed in perturbation
theory, and is thus of nonperturbative origin. From the kinematic
constraints in eq.~(\ref{e6}) and the on-shell condition of the
$u$-quark jet, the lower limit of $z$ should be
$z_{\rm min}=x $, instead of $z_{\rm min}=0 $.

The tree-level expressions for the convolution factors $J$ and $H$ are
given by
\begin{eqnarray}
\ J^{(0)}&=&\delta(P_u^2)=
\delta \left(M^2_B \left[1-y_0+y-(1-z)(1-\frac{y}{x})-\frac{2 {\bf p_\bot}
\cdot{\bf p_{\nu\bot}} }{M^2_B} -\frac{{\bf p_{\bot}}^2 }{M^2_B}\right]
\right)\;,
\nonumber\\
\ H^{(0)}&\propto&(P_b\cdot p_\nu)(p_\ell\cdot P_u)
=((P_B-p)\cdot p_\nu)(p_\ell\cdot P_u)
\nonumber \\
&\propto&(x-y)\left(y_0-x-(1-z)\frac{y}{x}+\frac{2 {\bf p_{\bot}}\cdot
{\bf p_{\nu\bot}} }{M^2_B}\right)\;.
\label{tl}
\end{eqnarray}
Equation (\ref{as}) can be regarded as an expression
at the intermediate stage in the derivation of conventional factorization
theorems. If the $\bf p_\bot$ dependence in $J$ and $H$ is negligible,
the variable $\bf p_\bot$ in $S$ can be integrated over, and eq.~(\ref{as})
reduces to the conventional factorization formula. However, it is obvious
from eq.~(\ref{tl}) that at least the $\bf p_\bot$ dependence in $J$ is not
negligible, especially in the end-point region. This is the reason
we introduce the transverse degrees of freedom into our analysis.

Suppose we consider higher-order corrections to eq.~(\ref{as}) from a gluon
crossing the final state cut, and route the loop momentum $\ell$ through,
say, the jet subprocess. Without losing generality, we approximate
$J(p^++\ell^+,P_u^-+\ell^-,{\bf
p_\bot}+ {\bf \ell_\bot})\approx J(p^+,P_u^-,{\bf p_\bot}+{\bf \ell_\bot})$
according to the kinimatic relations $\ell^+<p^+$, $\ell^-<P_u^-$ and
$\ell_\bot \approx {\bf p_\bot}$. Hence, the loop integral cannot be
performed unless the dependence of $J$ on transverse momentum is known.
This difficulty can be removed by Fourier transform,
\begin{equation}
J(p^+,P_u^-,{\bf p_\bot}+{\bf \ell_\bot})=\int\frac{d^2{\bf b}}{(2\pi)^2}
{\tilde J}(p^+,P_u^-,{\bf b})e^{i({\bf p_\bot}+{\bf \ell_\bot})
\cdot{\bf b}}\;,
\label{ft}
\end{equation}
where the impact parameter $b$ (Fourier conjugate variable of $p_\bot$)
measures the transverse distance travelled by the jet.
Using eq.~(\ref{ft}), the $\ell_\bot$ dependence is decoupled from the
jet function, and the factor $e^{i{\bf \ell_\bot}\cdot{\bf b}}$
is absorbed into the loop integral, which can then be performed.
Therefore, an extra factor $e^{i\ell_\bot\cdot {\bf b}}$ is associated
with each gluon crossing the final state cut in our formalism.

To further simplify the factorization formula, we neglect those terms
involving ${\bf p_{\nu\bot}}$ in $J$ and $H$.
This is a good approximation
for $x\to 0$ and $x\to 1$, since for $x\to 0$ contributions from
transverse momenta are not important, and for $x\to 1$ the magnitude
${p_{\nu\bot}}= M_B\sqrt{(y_0-x-y/x)y/x}$ vanishes.
Equation~(\ref{as}) then becomes
\begin{eqnarray}
\frac{1}{\Gamma^{(0)}_\ell}\frac{d^3\Gamma}{dxdydy_0}
&=&M_B^2\int^{1}_x{dz} \int \frac{d^2{\bf b}}{(2\pi)^2}
\nonumber \\
&&\times
{\tilde S}(z,{\bf b},\mu){\tilde J}(z,P_u^-,{\bf b},\mu)
H(z,P_{u}^-,\mu)\;.
\label{asb}
\end{eqnarray}

\section{Resumming the Jet, Soft and Hard Subprocesses}

It can be shown that
the dominant subprocesses $J$, $S$ and $H$ contain large logarithms form
radiative corrections. In particular, $J$ gives rise to double (leading)
logarithms in the end point region. These large corrections spoil the
perturbation theory and must be organized.
In what follows we shall demonstrate in details how to
resum these large corrections up to next-to-leading logarithms. The first
step in resummation is to map out the leading regions of radiative
corrections. We work in axial gauge $n \cdot A =0$, where $n$ is a
space-like vector. The ${\cal O}(\alpha_s)$ diagrams that contain large
double logarithms in axial gauge at the end point are figs.~2a and 2b.
The self-energy diagram fig.~2c and the diagram
with a soft gluon connecting the two heavy quark lines, as shown in fig.~2d,
give only single soft logarithms.

In the collinear region with the loop momentum $\ell$ parallel to $P_u$
and in the soft region with $\ell\to 0$ we can eikonalize the heavy
b-quark line. Then figs.~2a and 2b are factorized out of the cross section,
and they are the diagrams that $J$ absorbs, as shown in fig.~3.
With the eikonal approximation, the b-quark propagator is expressed as
$1/v\cdot \ell$ to the order $1/M_B$ with $v=(1,1,{\bf 0_\bot})$. Hence,
the factorization in eq.~(\ref{asb}) is in fact valid to ${\cal O}(1/M_B)$.
The physics involved in this approximation is that a soft gluon or a gluon
moving parallel to $P_u$ can not explore the details of the $b$ quark, and
its dynamics can be factorized. This is consistent with the HQEFT, where the
$b$ quark is treated as a classical relativistic particle carrying color
source. Since the large mass $M_B$ does not appear in the eikonal
propragator, the only large scale in $J$ is $P_u^-$.

The remaining diagrams figs.~2c and 2d that give single soft logarithms
are grouped into $S$. It is then obvious that $S$ depends only on the
properties of the bound state $|B\rangle$, but not on the particular short
distance subprocess. Therefore, $S$ is a universal function describing the
distribution of the $b$ quark inside the $B$ meson. In fact, in the $1/M_b$
limit one can identify $S$ as the model independent ``shape function" or
``primodial function" obtained from the resummation in \cite{N,neubert} and
\cite{bsuv}, respectively.

The basic idea of the resummation technique is as follows. If the double
logarithms are organized into an exponential form, ${\tilde J}\sim
\exp[-\ln P_u^-\ln(\ln P_u^-/\ln b)]$, one can simplify the analysis
by studying the derivative of $\tilde J$, $d{\tilde J}/d\ln P_u^-=C
{\tilde J}$ \cite{bb}.
Since the coefficient function $C$ contains only single logarithms, it can
be treated by renormalization group (RG) methods. In this way, one reduces
the complicated double-logarithm problem to a single-logarithm problem.

$\tilde J$ is scale invariant in the gauge
vector $n$ as shown by the gluon propagator in axial gauge,
$\frac{-i N^{\mu\nu}(\ell)}{\ell^2+i\epsilon}$, with
\begin{equation}
N^{\mu\nu}(\ell)=
\ g^{\mu\nu}-\frac{n^\mu\ell^\nu+\ell^\mu n^\nu}{n\cdot \ell}
\ + n^2\frac{\ell^\mu\ell^\nu}{(n\cdot \ell)^2} \;.
\label{e13}
\end{equation}
Therefore, $\tilde J$ must depend on $P_u^-$ through the ratio
$(P_u\cdot n)^2/n^2$. It is then easy to show that the differential operator
$d/d\ln P_u^-$ can be replaced by $d/dn$ using a chain rule,
\begin{equation}
\ \frac{d\tilde J}{d\ln P^-_u}=-\frac{n^2}{v'\cdot n}v'^\alpha
\ \frac{d}{dn^\alpha}{\tilde J}
\end{equation}
with the vector $v'=(0,1,{\bf 0_\bot})$. This simplifies the task
tremendously, because the momentum $P_u$ flows through both quark and gluon
lines, but $n$ appears only in gluon propagators.

Applying $d/dn$ to the gluon propagator, we obtain
\begin{equation}
\ \frac{d}{dn_\alpha}N^{\mu\nu}=-\frac{1}{\ell\cdot n}(N^{\mu\alpha}\ell^\nu
\                      +N^{\nu\alpha}\ell^\mu)\;.
\end{equation}
The momentum $\ell$ that appears at both ends of the differentiated gluon
line is contracted with the vertex, where the gluon attaches the $u$ quark
or the eikonal lines. Next we add up all diagrams with different
differentiated gluon propagators, and apply the Ward identity. Finally, we
arrive at a differential equation for $\tilde J$ as shown in fig.~4,
where the square vertex represents
\begin{equation}
\ gT^a\frac{n^2}{(v'\cdot n)(\ell\cdot n)}v'^\alpha\;,
\end{equation}
with $T^a$ the color matrix. The factor 2 counts the two external quark
lines of $\tilde J$. An important feature of the
square vertex is that the gluon momentum $\ell$ does not give rise to
collinear divergences because of the nonvanishing $n^2$. The
leading regions of $\ell$ are then soft and ultraviolet, in which
the subdiagram containing the square vertex can be factorized as shown in
fig.~5 at ${\cal O}(\alpha_s)$. Hence, the differentiation really turns the
double-logarithm problem into a single-logarithm problem as stated before.

To seperate the soft and ultraviolet scales in fig.~5,
we introduce a function ${\cal K}$ to organize the soft
logarithms from the four diagrams figs.~5a-5d, and a function
${\cal G}$ for the ultraviolet divergences from figs.~5e and 5f.
Double counting is avoided by the subtraction in ${\cal G}$.
Generalizing ${\cal K}$ and ${\cal G}$ to all orders, we derive the
differential equation for $\tilde J$,
\begin{equation}
\ \frac{d}{d\ln P^-_u}{\tilde J}(P^-_u,b,\mu)=
\ 2\left[ {\cal K}(b\mu,\alpha_s(\mu))+{\cal G}(P_u^-/\mu,
\alpha_s(\mu))\right] \ {\tilde J}(P^-_u,b,\mu)\;.
\label{ni}
\end{equation}
The scale $b$ in $\cal K$ arises from fig.~5b, which contains an extra
factor $e^{i{\bf \ell_\bot \cdot b}}$ as explained in Section 2.

At the one-loop level, the renormalized function ${\cal K}$
calculated in dimensional regularization for the gauge vector
$n\propto (1,-1,{\bf 0_\bot})$ is given by,
\begin{eqnarray}
 {\cal K}&=&{\rm fig.~5a+fig.~5b+fig.~5c+fig.~5d}-\delta {\cal K}\;,
\end{eqnarray}
with
\begin{eqnarray}
{\rm fig.~5a+fig.~5b}&=& -g^2C_F\mu^\epsilon\int \frac{d^{4-\epsilon}\ell}
{(2\pi)^{4-\epsilon}}
\nonumber \\
&& \times \left\{\pi\delta(\ell^2) e^{i{\bf \ell_\bot\cdot b}}-\frac{i}
{\ell^2+i\epsilon}\right\}
\frac{n^2v'_\alpha v'_\beta}{(v'\cdot n)(\ell\cdot n)(\ell\cdot v')}
N^{\alpha\beta}(\ell)
\nonumber \\
& =&-\frac{\alpha_s}{2\pi}C_F\left\{\frac{2}{\epsilon}+{\rm ln}\pi\mu^2
b^2 e^\gamma\right\}\;,
\end{eqnarray}
\begin{eqnarray}
{\rm fig.~5c+fig.~5d}&=& -g^2C_F\mu^\epsilon\int \frac{d^{4-\epsilon}\ell}
{(2\pi)^{4-\epsilon}}
\nonumber \\
&& \times \left\{\pi\delta(\ell^2) e^{i{\bf \ell_\bot}\cdot {\bf b}}
-\frac{i}{\ell^2+i\epsilon}\right\}
\frac{n^2v'_\alpha v_{\beta}}{(v'\cdot n)(\ell\cdot n)(\ell\cdot v)}
N^{\alpha\beta}(\ell)
\nonumber \\
& =& 0\;.
\end{eqnarray}
$\delta\cal K$ is the scheme-dependent counterterm which will be specified
later. $C_F=4/3$ is
the color factor, and $\gamma$ is the Euler constant. Similarly, $\cal G$
is given by
\begin{equation}
{\cal G}={\rm fig.~5e+fig.~5f}-\delta{\cal G}\;,
\end{equation}
with
\begin{eqnarray}
{\rm fig.~5e+fig.~5f}
&=& -g^2C_F\mu^\epsilon\int \frac{d^{4-\epsilon}\ell}{(2\pi)^{4-\epsilon}}
\frac{\not\! P_u-\not\!\ell}{(P_u-\ell)^2+i\epsilon}
\nonumber \\
&& \times \left\{\gamma_\alpha +\frac{\not\! P_u-\not\! \ell}
{\ell\cdot v'}v'_\alpha\right\}
\frac{n^2 v'_\beta}{(v'\cdot n)(\ell\cdot n)}
\frac{N^{\alpha\beta}(\ell)}{\ell^2+i\epsilon}
\nonumber \\
& =&-\frac{\alpha_s}{2\pi}C_F\left\{-\frac{2}{\epsilon}+{\rm ln}
\frac{(P_u^-)^2\nu e^{\gamma-1}}{\pi \mu^2}\right\}\;,
\end{eqnarray}
and $\nu =(n\cdot v')^2/|n^2|$ is the gauge factor.

In the $\overline{\rm MS}$ scheme with
\begin{eqnarray}
{\delta \cal K}=-\delta {\cal G}=
-\frac{\alpha_s}{2\pi}C_F\left(\frac{2}{\epsilon}+\ln 4\pi e^{-\gamma}
\right)\;,
\label{e22}
\end{eqnarray}
the one-loop ${\cal K}$ and ${\cal G}$ are given by
\begin{eqnarray}
{\cal K}&=&-\frac{\alpha_s}{2\pi}C_F\ln\left(\frac{b^2\mu^2
e^{2\gamma}}{4}\right)\;,
\nonumber \\
{\cal G}& =&-\frac{\alpha_s}{2\pi}C_F\ln\left(
\frac{4(P_u^-)^2\nu}{\mu^2 e}\right)\;.
\end{eqnarray}
We have seperated the two scales in $\tilde J$, $b$ and $P_u^-$, into the
functions ${\cal K}$ and ${\cal G}$, respectively, such that RG methods are
applicable to the summation of the corresponding single logarithms.
Although ${\cal K}$ and ${\cal G}$ possess individual ultraviolet
pole, their sum is finite, and thus a RG invariant quantity. The RG
equations for the renormalized ${\cal K}$ and ${\cal G}$ are
\begin{eqnarray}
\mu\frac{d}{d\mu}{\cal K}(b\mu ,\alpha_s(\mu))=-\gamma_{\cal K}
(\alpha_s(\mu))=-\mu\frac{d}{d\mu}{\cal G}(P_u^-/\mu,\alpha_s(\mu))
\label{asd}
\end{eqnarray}
with
\begin{equation}
\gamma_{\cal K}=\mu\frac{d}{d\mu}\delta{\cal K}=
-\mu\frac{d}{d\mu}\delta{\cal G}\;,
\end{equation}
the anomalous dimension of ${\cal K}$. To two loops, $\gamma_{\cal K}$
is given by \cite{BS}
\begin{equation}
\ \gamma_{\cal K}=\frac{\alpha_s}{\pi}C_F+\left(\frac{\alpha_s}{\pi}
\right)^2C_F\left[C_A\left(\frac{67}{36}-\frac{\pi^2}{12}\right)-\frac{5}
{18}n_f\right]\;,
\end{equation}
where $n_{f}=4$ is the number of quark flavors and $C_A=3$ is the
color factor.

Equation~(\ref{asd}) has the solution
\begin{eqnarray}
{\cal K}(b\mu ,\alpha_s(\mu))+{\cal G}(P_u^-/\mu,\alpha_s(\mu))
&=&{\cal K}(1,\alpha_s(P_u^-))+{\cal G}(1,\alpha_s(P_u^-))
\nonumber \\
&& -\int^{P_u^-}_{1/b}\frac{d\bar\mu}{\bar\mu}
A(\alpha_s(\bar\mu))
\label{kg}
\end{eqnarray}
with the anomalous dimension
\begin{equation}
A(\alpha_s)=\gamma_{\cal K}(\alpha_s)+\beta(g)\frac{\partial}{\partial g}
{\cal K}(1,\alpha_s)\;.
\end{equation}
Substituting eq.~(\ref{kg}) into (\ref{ni}), we
obtain the evolution of $\tilde J$ in $P^-_u$ and $b$,
\begin{equation}
\ {\tilde J}(P_u^-,b,\mu )=\exp[-2s(P_u^-,b)]{\tilde J}(b,\mu )\;.
\end{equation}
The RG invariant Sudakov exponent is given by \cite{LS}
\begin{eqnarray}
s(P_u^-,b)&=&\frac{A^{(1)}}{2\beta_{1}}\hat{q}\ln\left(\frac{\hat{q}}
{\hat{b}}\right)+
\frac{A^{(2)}}{4\beta_{1}^{2}}\left(\frac{\hat{q}}{\hat{b}}-1\right)-
\frac{A^{(1)}}{2\beta_{1}}\left(\hat{q}-\hat{b}\right)
\nonumber \\
&& -\frac{A^{(1)}\beta_{2}}{4\beta_{1}^{3}}\hat{q}
\left[\frac{\ln(2\hat{b})+1}
{\hat{b}}-\frac{\ln(2\hat{q})+1}{\hat{q}}\right]
\nonumber \\
&& -\left[\frac{A^{(2)}}{4\beta_{1}^{2}}-\frac{A^{(1)}}{4\beta_{1}}
\ln\left(\frac{e^{2\gamma-1}}{2}\right)\right]
\ln\left(\frac{\hat{q}}{\hat{b}}\right)
\nonumber \\
&& +\frac{A^{(1)}\beta_{2}}{8\beta_{1}^{3}}\left[
\ln^{2}(2\hat{q})-\ln^{2}(2\hat{b})\right]\;,
\label{sss}
\end{eqnarray}
with the variables
\begin{equation}
\ \hat q \equiv \ln\left(\frac{P_u^-}{\Lambda}\right)\;,
\ \hat b \equiv \ln\left(\frac{1}{b \Lambda}\right)\;.
\end{equation}
The QCD scale $\Lambda\equiv \Lambda_{\rm QCD}$ will be set to 0.2 GeV
in the numerical study in Section 5. The coefficients $\beta_{i}$ and
$A^{(i)}$ are
\begin{eqnarray}
\beta_{1}&=&\frac{33-2n_{f}}{12}\;,
\nonumber \\
\beta_{2}&=&\frac{153-19n_{f}}{24}\; ,
\nonumber \\
A^{(1)}&=&\frac{4}{3}\;,
\nonumber \\
A^{(2)}&=&\frac{67}{9}-\frac{\pi^{2}}{3}-\frac{10}{27}n_
{f}+\frac{8}{3}\beta_{1}\ln\left(\frac{e^{\gamma}}{2}\right)\; .
\label{12}
\end{eqnarray}

Having summed up the double logarithms, we concentrate on the single
logarithms in $\tilde S$, $H$ and the initial condition ${\tilde J}(b,\mu)$.
Since both the differential decay rate and the Sudakov exponent
$s(P_u^-,b)$ are RG invariant, we have the following RG equations:
\begin{eqnarray}
{\cal D}{\tilde J}(b,\mu)&=& -2\gamma_q {\tilde J}(b,\mu)\;,
\nonumber \\
{\cal D}{\tilde S}(b,\mu)&=& -\gamma_S{\tilde S}(b,\mu)\;,
\nonumber \\
{\cal D}H(P_u^-,\mu)&=& (2\gamma_q+\gamma_S)H(P_u^-,\mu)\;,
\label{ter}
\end{eqnarray}
with
\begin{equation}
\ {\cal D}=\mu\frac{\partial}{\partial\mu}+\beta(g)
\ \frac{\partial}{\partial g}\;.
\end{equation}
$\gamma_q=-\alpha_s/\pi$ is the quark anomalous dimension in axial
gauge, and $\gamma_S=-(\alpha_s/\pi)C_F$ is the anomalous dimension of
$\tilde S$. The function $\tilde S$ in fact contains soft single logarithms
only from fig.~2d, because the contribution from fig.~2c vanishes for
the guage vector $n\propto (1,-1,{\bf 0_\bot})$ under eikonal approximation.
Hence, $\gamma_S$ is derived from the evaluation of fig.~2d
straightforwardly.

Integrating eq.~(\ref{ter}), we obtain the evolution of all the convolution
factors,
\begin{eqnarray}
{\tilde J}(z,P_u^-,b,\mu)&=& {\rm exp}\left[-2s(P_u^-,b)-2\int^\mu_{1/b}
\frac{d{\bar\mu}}{{\bar\mu}}\gamma_q(\alpha_s(\bar\mu))\right]
{\tilde J}(z,b,1/b)\;,
\nonumber\\
{\tilde S}(z,b,\mu)&=& {\rm exp}\left[-\int^\mu_{1/b}
\frac{d{\bar\mu}}{{\bar\mu}}\gamma_S(\alpha_s(\bar\mu))\right]f(z,b,1/b)\;,
\nonumber\\
H(z,P_u^-,\mu)&=& {\rm exp}\left[-\int^{P_u^-}_\mu
\frac{d{\bar\mu}}{{\bar\mu}}[2\gamma_q(\alpha_s(\bar\mu))
+\gamma_S(\alpha_s(\bar\mu))]\right]H(z,P_u^-,P_u^-)\;.
\label{un}
\end{eqnarray}
We shall neglect the intrinsic $b$ dependence of the distribution function
$f$ below. If Sudakov suppression in the large-$b$ region is strong,
we may drop the evolution of $f$ and $\tilde J$ in $b$, which is proportional
to $\alpha_s(1/b)$. Hence, we assume $f(z,b,1/b)= f(z)$,
${\tilde J}(z,b,1/b)={\tilde J}^{(0)}(z,b)$, the Fourier
transform of the tree-level expression in eq.~(\ref{tl}), and
$H(z,P_u^-,P_u^-)=H^{(0)}(z,P_u^-)$.

Substituting eq.~(\ref{un}) into (\ref{asb}), we derive the
factorization formula for the inclusive semileptonic $B$ meson decay,
\begin{eqnarray}
\frac{1}{\Gamma^{(0)}_\ell}\frac{d^3\Gamma}{dxdydy_0}
&=&M_B^2\int^{1}_{x}
dz\int_0^{\infty}\frac{bdb}{2\pi}f(z){\tilde J}^{(0)}(z,b)
H^{(0)}(z,P_{u}^-)
\nonumber \\
&&\times
\exp[-S(P_u^-,b)]\;.
\label{as1}
\end{eqnarray}
The complete Sudakov exponent $S$ is given by
\begin{equation}
\ S(P_u^-,b)=2s(P_u^-,b)-\frac{5}{3\beta_1}{\rm ln}\frac{\hat P_u^-}
{\hat b}\;,
\label{qu}
\end{equation}
with $\hat P_u^-=\ln(P_u^-/\Lambda)$, which
combines all the exponents in eq.~(\ref{un}) and
includes both leading and next-to-leading logarithms.
It is straightforward to observe from eq.~(\ref{qu}) that the Sudakov
form factor $e^{-S}$ falls off quickly at large $b\sim 1/
\Lambda$, where $\alpha_s(1/b) > 1$ and perturbation theory fails.
Hence, the Sudakov form factor guarantees that  main contributions to the
factorization formula come from the small $b$, or short-distance, region,
and the perturbative treatment is indeed self-consistent.

We stress that our formalism is applicable to the entire range of the
spectrum, if the ${\bf p_{\nu\bot}}$ dependence in $J$ and $H$ was not
neglected, not only to the end-point region as in \cite{ks}. We neglected
the ${\bf p_{\nu\bot}}$ dependence for the sake of simplicity of
calculation. Since the Sudakov form factor is defined in the region
\begin{equation}
\left(1-\frac{y}{x}\right)\frac{M_B}{\sqrt{2}}=P_u^->\frac{1}{b}>\Lambda\;,
\end{equation}
there exist the suppression effects as $y<(1-\sqrt{2}\Lambda/M_B)x$.
The phase space with suppression expands to the largest extent for
$x\to 1$ and vanishes at $x\to 0$. Hence, lowest-order predictions receive
maximal suppression at the end point, corresponding to the presence of
large logarithms, but are modified only slightly at small $x$,
implying weaker logarithmic corrections.
This is consistent with our expectation for the Sudakov effects in
the whole range $0<x<1$.
\vskip 2.0cm

\section{Constructing the Universal Soft Function}
\hskip 0.6cm
Since all the double logarithmic corrections have been absorbed into the
jet subprocess, the soft function $\tilde S$ contains
only soft single logarithms from fig.~2d (fig.~2c does not contribute
because of the choice of the guage vector $n\propto (1,-1,{\bf 0_\bot})$).
These single logarithms can be summed by solving the RG equation
${\cal D}{\tilde S}=-\gamma_S {\tilde S}$
as shown in eq.~(\ref{ter}). The solution has been given in eq.~(\ref{un}),
\begin{eqnarray}
{\tilde S}(z,b,\mu)= {\rm exp}\left[-\int^\mu_{1/b}
\frac{d{\bar\mu}}{{\bar\mu}}\gamma_S(\alpha_s(\bar\mu))\right]f(z)\;,
\label{de}
\end{eqnarray}
where the initial condition $f(z)$ for the evolution of $\tilde S$ must be
determined phenomenologically. From the definition of
$\tilde S$, it is obvious that $f$ depends only on the properties of the bound
state $|B\rangle$, but not on the particular short distance subprocess.
Therefore, $f$ is a process-independent universal function describing the
distribution of the $b$ quark inside a $B$ meson.

As mentioned in the previous section, in the $1/M_b$ limit one
can identify $f$ as the model independent shape function or primodial
function obtained from the resummation of OPE in
Refs.~\cite{N} and \cite{bsuv,neubert}. It was observed by Korchemsky and
Sterman \cite{ks} that one can deduce some nonperturbative
information of $f$ from the perturbative resummation of soft gluons.
Theses authors argued that the infrared renormalons appearing during the
resummation procedure produce power-correction ambiguities.
Hence, there should exist corresponding nonperturbative power-correction
ambiguities to render the physical $f$ well defined.
To leading order, with a ``minimal" ansatz for integrating the first
infrared renormalon singularity, Korchemsky and Sterman  \cite{ks} derived
\begin{equation}
\ f_{KS}(z)=\frac{1}{\sqrt{2\pi\sigma^2}}{\rm exp}
\ \left[-\frac{(1-z)^2}{2\sigma^2}\right]\;,
\label{qqq}
\end{equation}
which describes a Gaussian distribution with width $\sigma$ around
$z=1$. Requiring that the minimal ansatz be
consistent with HQEFT and OPE, one fixes the width to be
\begin{equation}
\ \sigma^2=\frac{\mu^2_\pi}{3M_b^2}\;,
\end{equation}
where $\mu^2_\pi=0.54\pm 0.12$ GeV$^2$ \cite{BB} is obtained from QCD
sum rule estimation.

Recall that $f_{KS}$ was derived by integrating only the first infrared
renormalon. This fact is reflected in the nonvanishing property of
$f_{KS}$ in the unphysical region $z>1$, and in its validity in the
vicinity of $z=1$. However, comparing $f_{KS}$ with the phenomenological
distribution function \cite{pssz},
\begin{equation}
\ f_{P}(z)=N_P\frac{z(1-z)^2}{[(1-z)^2+\epsilon_Pz]^2}\;,
\label{tres}
\end{equation}
one finds that they do share the same leading $(1-z)^2$ behavior.
$f_P$ was obtained from the experiments
of the $B$ meson production in $e^+e^-$ annihilation by applying a
crossing to the light quark and then a time reversal transformation
\cite{pssz}. The constant $N_P=0.133068$ is the normalization, and
$\epsilon_P=0.006$ is the shape parameter.
Inspired by this observation, we postulate the following two-parameter
distribution function for the $B$ meson,
\begin{equation}
\ f_B(z)=N\frac{z(1-z)^2}{[(z-a)^2+\epsilon z]^2}\theta (1-z)\;,
\label{qun}
\end{equation}
which has the correct leading $(1-z)^2$ behavior near $z=1$. The purpose of
introducing one more parameter $a$ in eq.~(\ref{qun}) is to allow a
consistency check on $f_P$ in eq.~(\ref{tres}).

We now relate the parameters $a$ and $\epsilon$ to the
hadronic matrix elements of some local operators derived from
HQEFT based OPE, which are standard techniques \cite{N,dsu,neubert}.
The distribution function in axial gauge is defined by
\begin{eqnarray}
f_B(z)&=&\frac{1}{2\sqrt{2}}\int \frac{dy^-}{2\pi} e^{i (1-z)P_B^+y^-}
\langle B|{\bar b_v}(0)b_v(y^-)|B\rangle\;,
\label{hui}
\end{eqnarray}
in which the large momentum of the rescaled heavy quark field $b_v(x)$
has been projected out as usual in HQEFT by
\beq
b_v(x)=e^{iM_b v\cdot x}b(x)\;.
\eeq
Hence, $f_B(z)$ is independent of the $b$-quark mass as it is written in
terms of a matrix element in eq.~(\ref{hui}). Note that the Dirac matrix
structure has been properly factorized as shown in eq.~(\ref{hui}).

To connect with  HQEFT, we write the $b$ quark momentum as $P_b=P_B-p=M_bv+
(M_B-M_b)v-p$, and identify the residual momentum of the $b$ quark as
$k={\bar \Lambda}v-p$, ${\bar \Lambda}=M_B-M_b$ being the effective mass of
the light partons in the $B$ meson. The probability to find a $b$ quark with
light-cone residual momentum $k^+$ inside the $B$ meson, $f_r(k^+)$, has
been defined in \cite{neubert}. The moments of $f_r(k^+)$,
\begin{equation}
A_n=\int k^{+n}f_r(k^+) dk^+,
\label{mom}
\end{equation}
have been derived in HQEFT based OPE as \cite{neubert}
\begin{equation}
A_0=1\;,\;\;\;\; A_1=0\;,\;\;\;\; A_2=\frac{M_B^2}{3}K_b\;,...
\end{equation}
These moments are expressed in terms of hadronic matrix elements
corresponding to the structure of the $1/M_b$ expansion, with
\beq
\ K_b\equiv -\frac{1}{2 M_B}\langle B|\bar b_v\frac{(iD)^2}{2M_b^2}b_v|B
\rangle\;.
\label{eq49}
\eeq
The vanishing $ {\cal O}(\Lambda_{\rm QCD}/{M_B})$ contribution to the first
moment $A_1$ is consistent with the conclusion from the renormalon
analysis that the first nontrivial power correction begins at
${\cal O}(\Lambda^2_{\rm QCD}/M_B^2)$ and with our intuition for vanishing
average residual momentum of the $b$ quark inside the $B$ meson in the
heavy quark limit.

The moments of $f_B(z)$ can be expressed as local hadronic matrix elements
by performing an OPE of the bilocal operator $\bar b_v(0) b_v(y^-)$ in
eq.~(\ref{hui}) in the power of $1/M_b$. The relation between $f_B$ and
$f_r$ is then given by \cite{neubert}
\begin{equation}
f_B(z)dz=\left[f_r(k^+)+{\cal O} (1/M_b)\right] d k^+ \;,
\label{re}
\end{equation}
which reflects the difference of order ${\bar\Lambda}$ between the $B$
meson kinematics and the $b$ quark kinematics.
Using eq.~(\ref{mom}) and the definition $z=P_b^+/P_B^+$ or $k^+=P_B^+
(z-M_b/M_B)$, it is straightforward to derive the moments of $f_B$.
They are
\begin{eqnarray}
& &\int_0^1 f_B(z) dz=1\;,
\label{eq46} \\
& &\int^1_0 dz (1-z) f_B(z)={\bar \Lambda}/M_B +{\cal O}
(\Lambda^2_{\rm QCD}/M^2_B)\;,
\label{eq47}\\
& &\int^1_0 dz (1-z)^2 f_B(z)=\frac{\bar \Lambda^2}{M_B^2} +\frac{2}{3}
K_b+{\cal O} (\Lambda^3_{\rm QCD}/M^3_B)\;.
\label{eq48}
\end{eqnarray}
The first formula gives the correct normalization of $f_B$, which
corresponds to the total number of $b$ quarks inside a $B$ meson.
The second formula is related to the effective mass of light quarks,
${\bar \Lambda}$. The third formula gives the hadronic matrix element of
the kinematic operator, $K_b$.

To have a better insight, we examine if the distribution function
is consistent with our physical intuition for the behavior of the heavy
$b$ quark inside a $B$ meson. We calculate the mean $\mu$ and the
variance $\sigma^2$ of $f_B(z)$ from eqs.~(\ref{eq47}) and (\ref{eq48}),
and derive
\beq
\ \mu=1-\frac{\bar \Lambda}{M_B}+{\cal O}\left(\frac{\Lambda^2_{\rm
QCD}}{M_B^2} \right) \;,
\eeq
\beq
\ \sigma^2=\frac{2 K_b}{3} + {\cal O}\left(\frac{\Lambda^3_{\rm
QCD}}{M_B^3}\right) \;.
\eeq
Substituting the QCD sum rule $\cite{BB}$ and $B^*-B$ mass splitting
\cite{abc} results,
\beq
M_B=5.279 {\rm GeV}\;,\;\;\;\;M_b=4.776 {\rm GeV}\;,\;\;\;\; K_b=0.012\pm
0.0026\;,
\label{con}
\eeq
we obtain $\mu=0.90$ and $\sigma^2=0.0080\pm0.0017$, implying that
$f_B(z)$ peaks sharply around $z\approx \mu
\approx 1$ and has a width of ${\cal O}(\Lambda_{\rm QCD}/M_B)$.
The parameters $N$, $a$ and $\epsilon$ in $f_B(z)$ can be determined
using eqs~(\ref{eq46})-(\ref{eq48}) with (\ref{con}) inserted, which are
\begin{equation}
N=0.02609\;,\;\;\;\; a=0.9752\;,\;\;\;\; \epsilon=0.001699\;.
\label{eq55}
\end{equation}
The value of $a$, derived from the QCD constraints (by
taking finite number of moments only), is very close to unity. This
is consistent with the expectation from $f_P$. However, we emphasize that
$f_P$ is not quite consistent with HQEFT, its first and second moment
differing from eqs.~(\ref{eq47}) and (\ref{eq48}) by at least $45 \%$.

The distribution functions in eqs.~(\ref{tres}) and (\ref{qun})
will serve as the initial conditions of the soft function in (\ref{de}).
We then derive the PQCD factorization formula for the inclusive decay
$B\to X_u \ell\nu$, with the phenomenological inputs
satisfying the QCD constraints from HQEFT based OPE.

\section{The Charged Lepton Spectrum}
\hskip 0.6cm
In this section we evaluate eq.~(\ref{as1}) numerically for various
distribution functions. The charged lepton spectrum for the decay
$B\to X_u \ell\nu$ from the naive quark model is obtained by simply choosing
$f(z)=\delta (1-z)$ and ignoring the transverse momentum dependence in
$J^{(0)}$ and $H^{(0)}$. A simple calculation leads to
\beq
\frac{1}{{\Gamma_{\ell}}^{(0)}}\frac{d \Gamma}{dx}
=\frac{x^2}{6}\left(3-2x\right)\;,
\label{eq53}
\eeq
which corresponds to the solid curve (1) in fig.~6. This curve does not
fall off at the end point of the spectrum, contradicting
the observed behavior of the inclusive semileptonic decays of $B$ mesons.
The discrepancy implies that the tree-level analysis is not appropriate,
especially in the end-point region where PQCD corrections are important
as discussed in Section 3.

We then take into account Sudakov suppression from the resummation of
large radiative corrections. Substituting $f(z)=\delta (1-z)$,
$H^{(0)}=(x-y)(y_0-x)$ and the Fourier transform of $J^{(0)}=\delta(P_u^2)$
with $P_u^2=M_B^2(1-y_0+y-p_{\bot}^2/M_B^2)$ into eq.~(\ref{as1}),
we derive the modified quark model spectrum. This spectrum is,
after integrating eq.~(\ref{as1}) over $z$ and $y_0$, described by
\begin{eqnarray}
\frac{1}{{\Gamma_{\ell}}^{(0)}}\frac{d \Gamma}{d x}&=&M_B
\int_0^{x}dy \int_0^{1/\Lambda}db e^{-S
(P_u^-,b)}  (x-y) \eta
\nonumber \\
&& \times\left[(1+y-x) J_1(\eta M_B b)
-\frac{2}{M_B b} \eta J_2(\eta M_B b) +\eta^2 J_3 (\eta M_B b)\right]\;,
\label{e54a}
\end{eqnarray}
where $P_u^-=(1-y/x)M_B/\sqrt{2}$, $\eta=\sqrt{(x-y)(1/x-1)}$ and
$J_1$,$J_2$,$J_3$ are the Bessel functions of order 1, 2 and 3, respectively.
Note the presence of the Sudakov form factor $e^{-S}$ and the expression
in the square brackets which comes from ${\tilde J}^{(0)}$. The cutoff
$1/\Lambda$ of the impact parameter $b$ is set by the Sudakov form factor.
Numerical results of eq.~(\ref{e54a}) for $\Lambda=0.2$ GeV are shown by
the dashed curve (1) in fig.~6. Since we have neglected the
${\bf p_{\nu\bot}}$ dependence in $J$ and $H$ for simplicity,
eq.~(\ref{e54a}) is appropriate only for
small and large $x$. Therefore, to obtain the dashed curve (1), we evaluate
eq.~(\ref{e54a}) in the regions $0\le x\le 0.7$ and $0.9\le x \le 1$,
and then extrapolate from $x=0.7$ to 0.9 smoothly. The dependence on
$\Lambda$ in our analysis is also examined, and it is found that
predictions increase by only 10-20\% if $\Lambda$ was set to 0.1 GeV.

One observes immediately that the Sudakov effects alone are enough to
render the uprising free quark spectrum fall off at the end point.
This is consistent with our expectation that the inclusion of transverse
momenta and Sudakov suppression diminishes the on-shell configuration of
the outgoing $u$-quark jet. Another important feature in fig.~6 is that
the solid and dashed curves coincide with each other in the region
$x\to 0$. This indicates that the Sudakov effects
almost cease to contribute away from the end point as stated in Section 3.

The spectrum from the parton model without Sudakov suppression is obtained
by adopting $H^{(0)}=(x-y)[y_0-x-(1-z)y/x]$ and
$P_u^2=M_B^2[1-y_0+y-(1-z)(1-y/x)]$. With integration over $y_0$, we derive
\beq
\frac{1}{{\Gamma_{\ell}}^{(0)}}\frac{d \Gamma}{d x}=
\int_0^{x}dy \int_x^{1} dz f(z) (x-y)(y+z-x)\;,
\label{e55a}
\eeq
where $f(z)$ can be replaced by the distribution functions given in
eqs.~(\ref{tres}) and (\ref{qun}). Predictions from the use of $f_B$ and
$f_P$ are represented by the solid curves (2) and (3) in fig.~6,
respectively. Both of the spectra deviate from the quark model one slightly
at small $x$, and vanish at the end point. Since eq.~(\ref{e55a})
incorporates nonperturbative effects from primodial heavy quark motion
\cite{bsuv,neubert} (or soft dynamics in our formalism) through $f$, we
conclude that these nonperturbative corrections are indeed important in
the end-point region.

At last, including Sudakov suppression into eq.~(\ref{e55a}),
we arrive at the charged lepton spectrum of the $B\to X_u\ell\nu$ decay
that takes into account both large perturbative and nonperturbative
corrections,
\begin{eqnarray}
\frac{1}{{\Gamma_{\ell}}^{(0)}}\frac{d \Gamma}{d x}&=&{M_B}
\int_0^{x}dy \int_0^{1/\Lambda}db \int_x^{1} dz f(z)(x-y) \xi
\biggl[(z+y-x) J_1 (\xi M_B b)
\nonumber \\
&&
\left.-\frac{2}{M_B b} \xi J_2(\xi M_B b)
+\xi^2 J_3(\xi M_B b)\right] e^{-S (P_u^-,b)}\;,
\label{e54}
\end{eqnarray}
with $\xi=\sqrt{(x-y)(z/x-1)}$. Predictions from $f_B$ and $f_P$ are shown by
the dashed curves (2) and (3), respectively. They coincide with the solid
curves at small $x$, but descend by about 50\% at $x\to 1$ as shown in
fig.~7, implying strong suppression in the end-point region.
The slope of the spectrum then becomes smoother as expected.

{}From fig.~6 we evaluate the total decay rate $\Gamma/\Gamma^{(0)}_\ell$,
and results along with the Sudakove effects are displayed in Table. 1.
We find that the overall suppression from the Sudakov effects is 8\% for
the quark model and less than 7\% for the use of $f_B$ and $f_P$.
The two distribution functions lead to about $20 \%$ difference in the
total decay width. Comparing to the drastic distinction between $f_B$ and
$f_P$ with $N_P/N\sim 5$ and $\epsilon_P/\epsilon\sim 4$, our formalism
is quite insensitive to the choice of different distribution functions.
The overall Sudakov suppression of less than $8 \%$ indicates that PQCD
corrections are actually not important for most part
of the spectrum. This is consistent with the fact that corrections from
transverse momenta are an ${\cal O}(1/M_Q^2)$ effect.

Note the 30\% suppression on the quark-model results from the distribution
function $f_B$ (the suppression from $f_P$ is even stronger). It
has been found in HQEFT that effects from nonperturbative corrections are
only of ${\cal O}(1/M_B^2)$, which should be less than 5\% (see \cite{neu}
and references therein). The small nonperturbative corrections to the total
decay rate are closely
related to the vanishing first moment of the residual momentum structure
function $f_r$. This apparent discrepancy can be traced back to the fact
that the $B$ meson kinematics is employed in our formalism, while the
$b$-quark kinematics is employed in the conventional approaches. Hence, in
our quark-model analysis the $b$ quark in fact carries the full momentum
$P_b=P_B$, and thus the charged lepton energy $E_\ell$ can reach the maximum
$M_B/2$ $(x=1)$. This momentum configuration is allowed in factorization
theorems if transverse degrees of freedom of partons were included, because
its invariant $P_b^2=M_B^2-{\bf k}_T^2$ may still be close to the mass shell
in the region without Sudakov suppression. Without ${\bf k}_T$,
eq.~(\ref{eq53}) should be regarded as an expression that is generated in
our formalism to bear the same form as the leading-power results in HQEFT.
For a free $b$ quark with momentum $M_bv$, $E_\ell$ can only reach $M_b/2$,
instead of $M_B/2$. Strickly speaking, our quark-model predictions and
the leading-power predictions in HQEFT have different meanings.

We stress that our results do not violate the conclusion from HQEFT, if
they were interpreted in a proper way. To confirm this, we
identify $P_b=(M_b^2/2P_B^-,P_B^-,{\bf 0})$ as the momentum carried
by a free $b$ quark in factorization theorems for $B$ meson decays, where
the minus component $P_b^-$ has been set to $P_B^-$. That is, the free $b$
quark is not at rest inside the $B$ meson. We then
reexpress eq.~(\ref{e55a}) into a form similar to that in \cite{neu}:
\beq
\frac{1}{{\Gamma_{\ell}}^{(0)}}\frac{d \Gamma}{d x}=
F(x)\theta\left(\frac{M_b^2}{M_B^2}-x\right)+F\left(\frac{M_b^2}{M_B^2}
\right)M(x)\;,
\label{ned}
\eeq
with $F(x)=x^2(3-2x)/6$ being the quark-model prediction derived from the
conventional approaches and
\begin{equation}
F\left(\frac{M_b^2}{M_B^2}\right)M(x)=
\int_0^{x}dy \int_x^{1} dz f(z) (x-y)(y+z-x)-
F(x)\theta\left(\frac{M_b^2}{M_B^2}-x\right)\;.
\label{nonp}
\end{equation}
The step function in eq.~(\ref{ned}) specifies the maximal $E_\ell$
in the decay of a free $b$ quark with the above momentum $P_b$. The
function $M(x)$, representing nonperturbative corrections to the $b$ quark
decay, coincides with the shape function $S(x)$ defined in \cite{neu}.

We shall show that the contribution from $M(x)$ to the total decay rate
is indeed of ${\cal O}(1/M_B^2)$. Integrating eq.~(\ref{nonp}) over $x$,
we obtain
\begin{equation}
F\left(\frac{M_b^2}{M_B^2}\right)\int_0^1 M(x)dx=\frac{1}{12}
\int_0^1 dzz^4f(z)-\int_0^{M_b^2/M_B^2} F(x)dx\;.
\label{nonp1}
\end{equation}
An arbitrary structure function $f$, which possesses the same moment as in
eq.~(\ref{eq47}), can be expanded in terms of $\delta$-functions:
\begin{equation}
f(z)=\delta(1-z)-\frac{\bar \Lambda}{M_B}\delta'(1-z)
+{\cal O}({\bar\Lambda}^2/M_B^2)\;.
\label{ex}
\end{equation}
Inserting eq.~(\ref{ex}) into (\ref{nonp1}), we justify straightforwardly
that the nonperturbative correction
\begin{eqnarray}
F\left(\frac{M_b}{M_B}\right)\int_0^1 M(x)dx&=&
\int_{M_b^2/M_B^2}^1 F(x)dx-\frac{1}{12}\frac{\bar\Lambda}{M_B}
\int_0^1 dz z^4\delta'(1-z)+{\cal O}({\bar\Lambda}^2/M_B^2)
\nonumber \\
&=&\frac{1}{3}\frac{\bar\Lambda}{M_B}-\frac{1}{3}\frac{\bar\Lambda}{M_B}
+{\cal O}({\bar\Lambda}^2/M_B^2)
\label{cancel}
\end{eqnarray}
vanishes at ${\cal O}(1/M_B)$ as concluded in \cite{neu}.
In summary, the quark-model contribution from the window between
$x=M_b^2/M_B^2$ and $x=1$, with a width of ${\cal O}(1/M_B)$, cancels the
${\cal O}(1/M_B)$ correction from the structure function, such that the
nonperturbative correction is of ${\cal O}(1/M_B^2)$.

According to eq.~(\ref{nonp1}), the suppression from nonperturbative
corrections is about 5\% for $f_B$ and more than 10\% for $f_P$. The
percentage for $f_P$ is still large, because its first moment does not
satisfy the requirement of HQEFT, and thus the cancellation at the power
$1/M_B$ is not complete.

The ambiguity from the choice of distribution functions can be removed,
once the spectrum of the decay $B\rightarrow X_s \gamma$ is available.
We can fix the universal $B$ meson distribution function
from these data, substitute the distribution function into our formula,
and predict the end-point spectrum of the decay $B\to X_u\ell\nu$.
A model independent extraction of the CKM matrix element $|V_{ub}|$
then becomes possible \cite{neubert,ks}.

\section{Conclusion}
\hskip 0.6cm
We have studied the inclusive semileptonic $B \rightarrow X_u\ell\nu$ decay
using the PQCD formalism. In order to seperate the $B\to X_u\ell\nu$ signals
from the $B\to X_c\ell\nu$ background, we must investigate the charged
lepton spectrum near the end-point region within an accuracy of about 330
MeV. It has been found that there exist  large perturbative corrections
in this region, which are resummed into the Sudakov form factor and included
into the factorization formula. The transverse degrees of
freedom of the $b$ quark diminish the on-shell configuration
of the outgoing $u$-quark jet. The quark-model spectrum then falls off at
the end point, consistent with the experimental observation. There is no
ambiguity associated with the kinematic gap, because we formulate
the factorization for $B$ meson, instead of $b$ quark, decays.

We have constructed a distribution function, whose parameters are
determined by the HQEFT based OPE, and whose width and mean are related to
hadronic matrix elements of the kinematic operator. These hadronic matrix
elements are then fixed by QCD sum rule results \cite{BB} and $B^*-B$
splitting data \cite{abc}. The distribution function, absorbing important
nonperturbative corrections from heavy quark Fermi
motion, can also render the quark model spectrum vanish at the end point.

We emphasize that our formalism incorporates both large perturbative and
nonperturbative corrections in the end-point region of inclusive $B$ meson
decays in a systematic way, and that it provides a natural normalization
for the spectra. This enables the direct extraction of the CKM matrix
element $|V_{ub}|$ from experimental data. When more data are available,
our formalisms can also be used to test PQCD in $B$ meson decays.

It is an important issue that current experimental data \cite{c21} of
the $B\to X \ell\nu$ branching ratio suggest
\beq
BR( B\to X\ell\nu) \leq 11 \%\;.
\label{e55}
\eeq
The naive quark-model prediction for this branching ratio is more than
$15 \%$. Although PQCD suppression at the end point is around $50 \%$,
the overall suppression amounts to $8 \%$ at most. With modification from
the massiveness of the charm quark, our formalism can be applied equally
well to the semileptonic decay $B\to X_c\ell\nu$. Hence, we conclude that
PQCD corrections suppress the overestimated theoretical value of the
semileptonic branching ratio only down to $13.8\%$ at best. On the other
hand, the distribution functions may decrease the quark-model
predictions by about 30\%. However, the distribution function is
universal as stressed before, and it is very plausible that it gives an
equal amount of suppression to nonleptonic decays. Therefore, introducing
a distribution function may not be able to remove the disagreement.

Based on the above discussion, we propose three possibilities to resolve
the discrepancy: (1) the distribution function suppresses semileptonic $B$
meson decays maximally, but does nonleptonic decays minimally. This may
arise from the different phase space in these two cases. (2) Factorization
theorems break down in $B$ meson decays. (3) New QCD effects or new
physics appears. Blok and Mannel \cite{BM} argued that factorization
theorems may still hold, and thus the confrontation
between data and theoretical predictions becomes acute.
To settle down the issue, a careful PQCD analysis of $B$ meson
nonleptonic decays is required. We shall discuss these subjects in a
forthcoming article.

\noindent
{\bf Acknowledgements:}
\vskip 0.3cm
We thank Dr. Xin-Heng Guo who participated in this work at the early stage.
We also thank T. Mannel for communicating their works on
local duality, and M. Shifman for useful discussion.
This work was supported by the National Science Council
of R.O.C. under Grant Nos. NSC85-2112-M194-009 and NSC85-2112-M-001-021.

\newpage
Table.1 The total decay rates for the quark model and for the use of
the distribution functions $f_B$ and $f_P$.
\vskip 1.0cm
\[\begin{array}{lccc} \hline\hline
\Gamma/\Gamma^{(0)}_\ell & \delta(1-x) & f_B & f_P \\ \hline
{\rm without}\;\; {\rm suppression}& 0.0833 & 0.0586 & 0.0446\\
{\rm with}\;\; {\rm suppression}& 0.0767 & 0.0548 & 0.0425 \\
{\rm Sudakov}\;\; {\rm effects}& 7.92\% & 6.48\% & 4.71\% \\
\hline\hline
\end{array} \]

\newpage
\begin{center}
{\large\bf Figure Captions}
\end{center}

\vskip 1.0cm
\noindent
{\bf Fig.1}:\\
Factorization of inclusive semileptonic decays of the $B$ meson
into a soft (S), a jet (J) and a hard (H) subprocess.
\vskip 1.0cm
\noindent
{\bf Fig.2}:\\
Lowest-order radiative corrections to the inclusive $B$ meson decays.
\vskip 1.0cm
\noindent
{\bf Fig.3}:\\
Factorization of the jet subprocess.
\vskip 1.0cm
\noindent
{\bf Fig.4}:\\
Graphic representation of eq.~(\ref{ni}).
\vskip 1.0cm
\noindent
{\bf Fig.5}:\\
Lowest-order diagrams for the functions ${\cal K}$ and
${\cal G}$.
\vskip 1.0cm
\noindent
{\bf Fig.6}:\\
Charged lepton spectra of the $B\to X_u\ell\nu$ decay for (1)
$f(z)=\delta (1-z)$, (2) $f(z)=f_B(z)$ and (3) $f(z)=f_P(z)$. The solid
(dashed) curves are derived without (with) Sudakov suppression.
\vskip 1.0cm
\noindent
{\bf Fig.7}:\\
Charged lepton spectra of the $B\to X_u\ell\nu$ decay near the end point
for the use of $f_B$ and $f_P$. Conventions are the same as those in fig.~6
but with the dotted curve corresponding to the dashed curve (3).


\newpage
\begin{thebibliography}{9}

\bibitem{aa}
M. Suzuki, \NP {\bf B145}, 420 (1978);\\
N. Cabibbo and L. Maiani, \PLB {\bf B79}, 109 (1978);\\
A. Ali and E. Pietrainen, \NP {\bf B154}, 519 (1979).
\bibitem{bb}
H-n. Li and H.L. Yu, \PRL {\bf 74}, 4388 (1995);\\
H-n. Li and H.L. Yu, \PLB {\bf B353}, 301 (1995).
\bibitem{cc}
B. Ong {\it et al.} ( CLEO Collaboration), \PRL {\bf 70}, 18 (1993).
\bibitem{af}
I. Bigi, B. Blok, M. Shifman and A. Vainshtein, \PLB {\bf B323},
408 (1994).
\bibitem{bf}
E. Bagan, P. Ball, V. Braun and P. Gosdzinsky, \PLB {\bf B342}, 362
(1995); {\it ibid.} \NP {\bf B432}, 3 (1994).\\
E. Bagan, P. Ball, B. Fiol, P. Gosdzinsky, \PLB {\bf B351}, 546 (1995).
\bibitem{ee}
J. Chay, H. Georgi and B. Grinstein, \PLB {\bf B247}, 399 (1990).
\bibitem{dd}
M. Shifman and M. Voloshin, \SNP {\bf 41}, 120 (1985);\\
B. Blok, L. Koyrakh, M. Shifman and A. Vainshtein, \PR {\bf D49}, 3356
(1993).
\bibitem{ff}
I.I. Bigi, M.A. Shifman, N.G. Uraltsev and A.I. Vainshtein, \PRL {\bf 71},
496 (1993).
\bibitem{gg}
I. Bigi {\it et al.} preprint CERN-TH-7132-94 (1994);\\
A. Falk, M. Wise and I. Dunietz, \PR {\bf D51}, 1183 (1995).
\bibitem{N}
M. Neubert, \PR {\bf D49}, 2472 (1994).
\bibitem{bsuv}
I.I. Bigi, M.A. Shifman, N.G. Uraltsev and A.I. Vainshtein, \IJMP {\bf A9},
2467 (1994).
\bibitem{dsu}
R.D. Dikeman, M. Shifman and N.G. Uraltsev, hep-ph/9505397.
\bibitem{neubert}
M. Neubert, Phys. Rev. D49, 4623 (1994).
\bibitem{kk}
A. Bareiss and E.A. Paschos, \NP {\bf B327}, 353 (1989).
\bibitem{jj}
G. Altarelli and S. Petrarca, \PLB {\bf B261}, 303 (1991);\\
G. Altarelli, N. Cabibbo, G. Corbo and L. Maiani, \NP {\bf B208}, 365 (1982).
\bibitem{ks}
G.P. Korchemsky and G. Sterman, \PLB {\bf B340}, 96 (1994).
\bibitem{BS}
J. Botts and G. Sterman, \NP {B325}, 62 (1989).
\bibitem{LS}
H.-n. Li and G. Sterman, \NP {B381}, 129 (1992).
\bibitem{BB}
P. Ball and V. Braun, \PR {\bf D49}, 2472 (1994).
\bibitem{pssz}
C. Peterson, D. Schlatter, I. Schmitt and P.M. Zerwas, \PR {\bf D27},
105 (1983).
\bibitem{abc}
Rev. of Particle Properties, \PR {\bf D50}, 1174 (1994).
\bibitem{neu}
M. Neubert, Phys. Rev. D49, 3392 (1994).
\bibitem{c21}
M. Athanas {\it et al.} (CLEO Collaboration), \PRL {\bf 73}, 3503 (1994).
\bibitem{BM}
B. Blok and T. Mannel, hep-ph/9505288.



\end{thebibliography}
\end{document}